\pdfoutput=1
\documentclass[%
 reprint,
nofootinbib,
 amsmath,amssymb,
 aps,
]{revtex4-2}

\usepackage{graphicx}
\usepackage{dcolumn}
\usepackage{bm}

\begin{document}

\title{Electronic Structure of Monolayer and Bilayer Black Phosphorus with Charged Defects}

\author{Martik Aghajanian}
\author{Arash A. Mostofi}
\author{Johannes Lischner}
\email{j.lischner@imperial.ac.uk}
\affiliation{Depts. of Physics and Materials and the Thomas Young Centre for Theory and Simulation of Materials, Imperial College London, London, SW7 2AZ, UK}

\date{\today}
\begin{abstract}
We use an atomistic approach to study the electronic properties of monolayer and bilayer black phosphorus in the vicinity of a charged defect. In particular, we combine screened defect potentials obtained from first-principles linear response theory with large-scale tight-binding simulations to calculate the wavefunctions and energies of bound acceptor and donor states. As a consequence of the anisotropic band structure, the defect states in these systems form distorted hydrogenic orbitals with a different ordering than in isotropic materials. For the monolayer, we study the dependence of the binding energies of charged adsorbates on the defect height and the dielectric constant of a substrate in an experimental setup. We also compare our results to an anisotropic effective mass model and find quantitative and qualitative differences when the charged defect is close to the black phosphorus or when the screening from the substrate is weak. For the bilayer, we compare results for charged adsorbates and charged intercalants and find that intercalants induce more prominent secondary peaks in the local density of states because they interact strongly with electronic states on both layers. These new insights can be directly tested in scanning tunneling spectroscopy measurements and enable a detailed understanding of the role of Coulomb impurities in electronic devices.
\end{abstract}

\maketitle
\section{\label{sec:intro} Introduction}

Black phosphorus (BP) is a semiconductor consisting of two-dimensional (2D) layers held together by van der Waals interactions. Its intrinsic properties, such as the direct band gap, a highly anisotropic electronic dispersion near the band extrema~\cite{Wu2015, Qiao} and high carrier mobilities~\cite{Rudenko2016,Liu2014}, are promising for applications in electronic and optoelectronic devices, including transistors, photodetectors and solar cells~\cite{Buscema2014, Lin2019, Long2016, Deng2014}. However, because of the rapid oxidation of BP in air, these devices typically require encapsulation to enhance their stability~\cite{Riffle2018,Wood2014}. Similar to graphite, individual monolayers can be exfoliated from bulk BP~\cite{Bagheri2016}, resulting in a 2D material commonly referred to as phosphorene. Monolayer and bilayer BP also exhibit an anisotropic dispersion relation as well as an enlarged band gap as a consequence of quantum confinement~\cite{Jing2016}. 
\newline

A detailed understanding of defects in monolayer and few-layer BP is crucial for improving devices. Defects can alter the material in a variety of ways, such as carrier doping, modifying carrier lifetimes and tuning the optical spectrum~\cite{Queisser945, Tamarat2006}. A variety of defects has been studied in few-layer BP, including intrinsic defects such as monovacancies or divacancies~\cite{Kiraly2017}, and extrinsic defects such as substitutional tin atoms or iodine and oxygen adatoms~\cite{Gaberle2018, Riffle2018}. If the defect donates electrons or holes to the BP, it becomes charged and induces a screened long-ranged Coulomb potential that results in the formation of bound defect states, which can be observed in scanning tunneling spectroscopy (STS) experiments. For example, Qiu and coworkers reported that $p$ dopants in BP give rise to lobed states with a spatial extent of several nanometers in the tunnelling spectrum~\cite{Qiu2017}. These lobed states were also observed by Wentink and coworkers~\cite{wentink2021substitutional}.  Additionally, devices that rely on 2D materials are typically placed on a substrate~\cite{Wood2014}, which can impact the properties of such defect states.
\newline

Valuable insights into the properties of charged defects in 2D materials can be obtained from simulations. However, modelling such defects is highly challenging because the slow decay of the screened defect potential requires the use of extremely large supercells containing thousands of atoms. As a consequence, standard first-principles techniques, such as density-functional theory (DFT), become unfeasible for these systems. To overcome these limitations of atomistic approaches, electronic continuum theories based on the effective mass approximation are often used~\cite{Rodin2014, Henriques2020}, which capture the long-range effects of the screened Coulomb interaction but are often not sufficiently accurate to predict the ordering of defect states or their binding energies. A favourable compromise between accuracy and computational efficiency is offered by tight-binding models. For example, we have recently used the tight-binding approach in conjunction with a screened defect potential obtained from first-principles linear response theory to study charged defects in graphene and also in monolayer transition metal dichalcogenides and found excellent agreement with experimental STS measurements~\cite{Aghajanian2018, Aghajanian2020, Corsetti2017,dos2018impact,wong2017spatially,lischner2019multiscale}. 
\newline

In this paper, we carry out atomistic tight-binding calculations to study the properties of charged defects in monolayer and bilayer BP. In the presence of a charged defect, bound states are formed with energies that lie in the band gap~\cite{Kohn1957, Bassani_1974}. We find that these in-gap states are localized to the defect, and form a lobed series with similar nodal structure as an anisotropic 2D hydrogen model. In particular, we calculate and study the defect wavefunctions and binding energies, as well as the local density of states which can be measured in STS experiments. We study donor and acceptor adsorbates in monolayer and bilayer BP, and also charged intercalants in bilayer BP. For the adsorbates, we investigate the dependence of the defect properties on the height of the defect above the BP as well as on the value of the substrate dielectric constant. We compare the results of the atomistic calculations to results obtained from a variational effective mass model to assess the performance of such widely used approaches, finding good agreement with the tight-binding results for large defect heights and strongly screening substrates. For bilayer BP, we compare the energies and wavefunctions of defect states bound to adsorbates and intercalants, finding similar energies of in-gap states but distinct behaviour of resonant states in the local density of states. Our study provides a detailed understanding of the properties of charged defects in monolayer and bilayer BP and their tunability through defect engineering, which will be important for the development of efficient devices.

\section{\label{sec:meth}Methods}

\begin{figure}[ht!]
    \centering
    \includegraphics[width=0.4\textwidth]{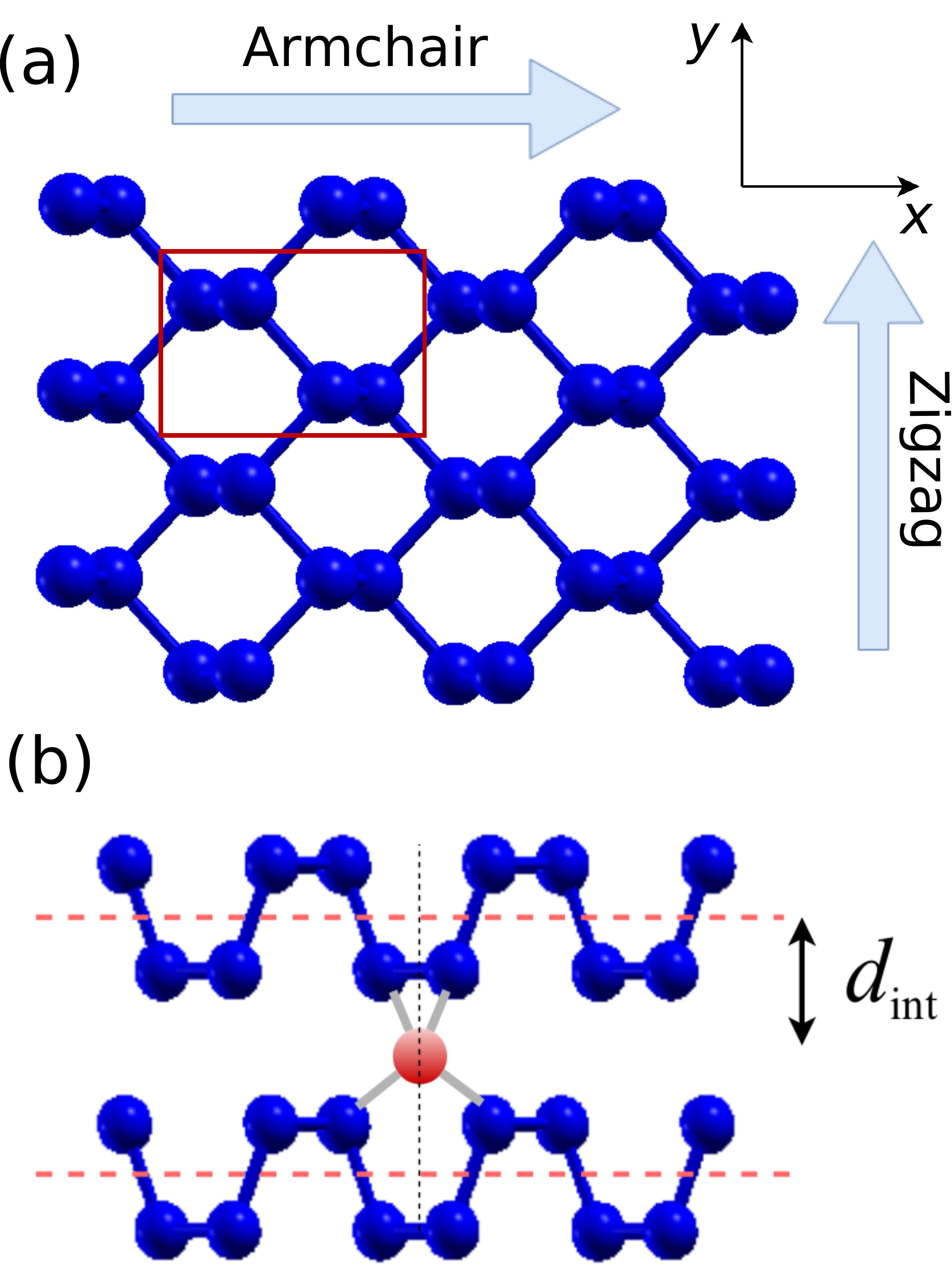}
    \caption{(a) Crystal structure of a black phosphorus monolayer. Armchair and zigzag directions are indicated and the unit cell of the pristine crystal is outlined in red. (b) Atomic structure of a charged intercalant in bilayer BP, where $d_{\text{int}}=2.72$~\AA\ is the distance between the intercalant and the top layer effective sheet (red dashed line).}
    \label{fig:structure}
\end{figure}

\subsection{Monolayer BP}

A top view of the atomic structure of the BP monolayer is shown in Fig.~\ref{fig:structure}(a). In contrast to graphene, the phosphorus atoms do not all lie in the same plane, but instead form two sublayers which are covalently bonded to each other. The charged adsorbate is located a distance $d$ directly above one of the atoms in the top sublayer. We consider both positively and negatively charged adsorbates with a charge of $\pm e$ ($e$ being the proton charge). Negatively charged adsorbates are also referred to as acceptor defects, while positively charged adsorbates are referred to as donors. Additionally, we assume that the monolayer is placed on a polarizable substrate with dielectric constant $\varepsilon_{\text{sub}}$. 

The total defect potential $V_\text{mono}^\pm$ experienced by an electron localized in one of the sublayers (with $V_\text{mono}^+$ referring to the top sublayer and $V_\text{mono}^-$ referring to the bottom one) can be expressed as the sum of the unscreened defect potential $V_{0}^{\pm}$ and the induced potential $V_{\text{ind}}^\pm$. The unscreened defect potential is given by
\begin{align}
     V_{0}^{\pm}(q) = \frac{v_q}{\varepsilon_{\text{bg}}}e^{-q(d + \frac{h_{\text{s}}}{2} \mp \frac{h_{\text{s}}}{2})},
     \label{eq:V0}
\end{align}
where $v_q=2\pi/q$ is the 2D Fourier transform of the Coulomb potential, $h_{\text{s}}=2.013$~\AA\ is the thickness of the monolayer defined as the distance between the two sublayers and $\epsilon_{\text{bg}}=(1+\epsilon_{\text{sub}})/2$ is the average dielectric constant of the substrate beneath and the vacuum above the monolayer.

\begin{figure}[ht!]
    \centering
    \includegraphics[width=0.49\textwidth]{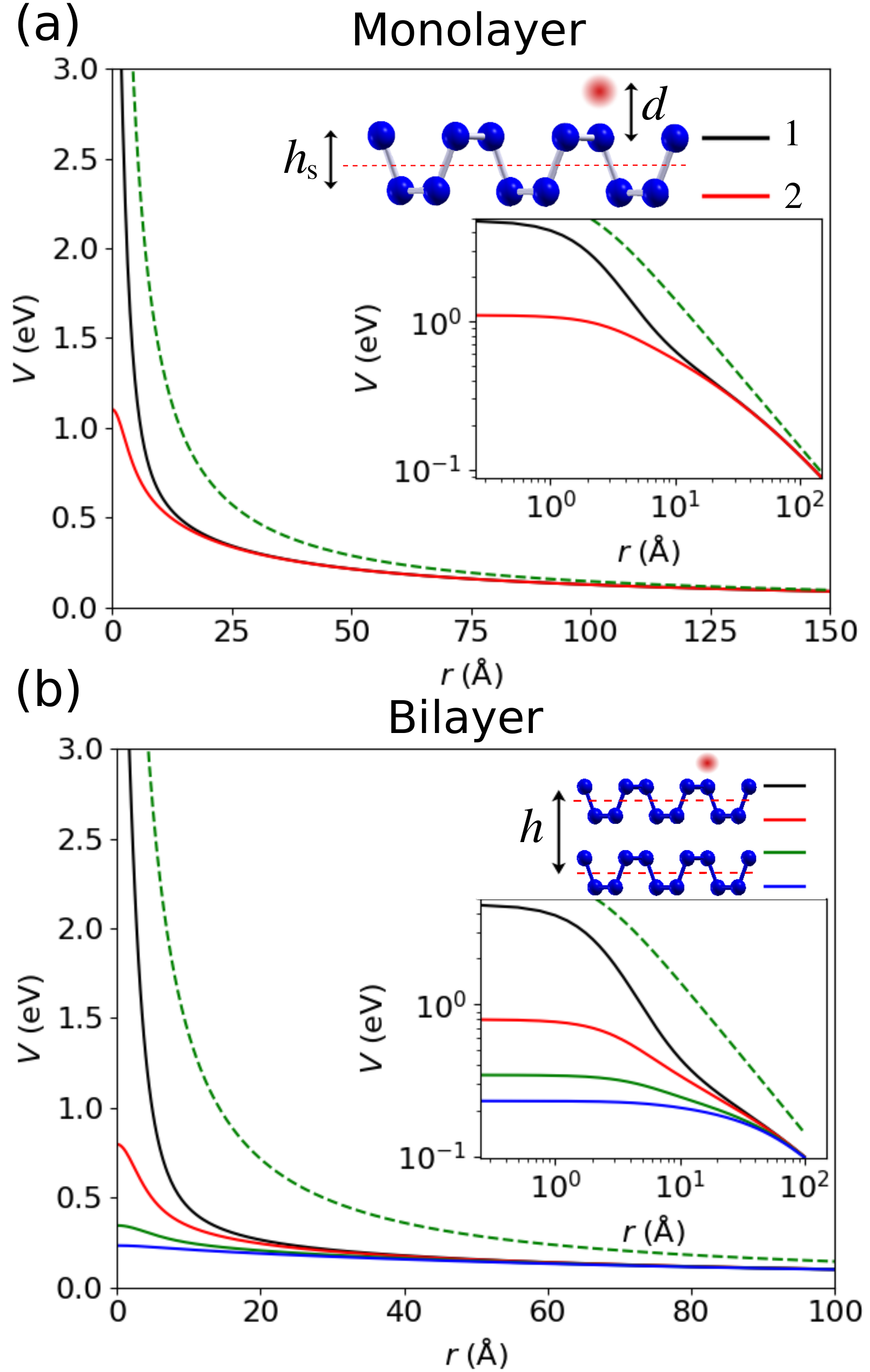}
    \caption{(a) Screened defect potential of a charged adsorbate on monolayer BP. The black line shows the potential in the top sublayer and the red line the result for the bottom sublayer. For comparison, the unscreened defect potential (green dashed line) is also shown. The adsorbate is located $d=2$~\AA\ above the top sublayer. The top inset shows a schematic of the system where the red dot indicates the adsorbate position and the red dashed line indicates the effective sheet where the induced charged is assumed to be localized. Also, $h_{\text{s}}$ is the monolayer thickness and the top and bottom sublayers are labelled as 1 and 2, respectively. (b) Sublayer-resolved screened potential of a charged adsorbate (located 2~\AA\ above the topmost sublayer) on bilayer BP. In the schematic, $h$ denotes the distance between the monolayers. The lower insets in (a) and (b) show log-log plots of the screened defect potentials.}
    \label{fig:mono_pot}
\end{figure}

The induced potential can be calculated from the induced electron density $\delta n_{\text{ind}}(q)$. To calculate the induced density, we employ the approach that was previously used for transition-metal dichalcogenides (TMDs)~\cite{Aghajanian2018,Aghajanian2020} and assume that the induced density is confined to an effective 2D sheet. In the case of TMDs, this effective sheet is located in the plane of the metal atoms. For monolayer BP, we assume the sheet to be located in the plane equidistant from the two sublayers, which is marked by the red dashed line in the inset of Fig.~\ref{fig:mono_pot}(a). Then the induced potential is the same for both sublayers and given by 
\begin{align}
    V^\pm_{\text{ind}}(q) = \frac{v_q}{\varepsilon_{\text{bg}}}e^{-qh_{\text{s}}/2}\delta n_{\text{ind}}(q).
    \label{eq:Vind}
\end{align}
Within linear response theory, the induced density can be expressed as 
\begin{align}
    \delta n_\text{ind}(q) = \chi(q) V_0(q),
\end{align}
where $V_0(q)=v_q e^{-q(d+\frac{h_{\text{s}}}{2})}/\varepsilon_{\text{bg}}$ is the unscreened defect potential in the effective sheet and $\chi(q)$ denotes the interacting response function which can be expressed in terms of the non-interacting response function $\chi_0(q)$ as
\begin{equation}
    \chi(q) = \frac{\chi_0(q)}{1 - v_q\chi_0/\varepsilon_{\text{bg}}},
\end{equation}
where $\chi_0(q)$ is obtained from first-principles calculations (see discussion below). Note that we have assumed isotropic screening in the above. This is justified by our explicit first-principles calculations and in agreement with recent experiments~\cite{Zhen2019}.

Combining Eqs.~\eqref{eq:V0} and \eqref{eq:Vind} yields the total defect potential
\begin{align}
    &V^{\pm}_{\text{mono}}(q)  = \\
    & \left[1 - \frac{v_q\chi_0(q)e^{-q(\frac{h_{\text{s}}}{2}\pm \frac{h_{\text{s}}}{2})}}{\varepsilon_{\text{bg}} - v_q
    \chi_0(q)(1 - e^{-q(\frac{h_{\text{s}}}{2}\pm \frac{h_{\text{s}}}{2})})}\right]^{-1} V_{0}^{\pm}(q).
\end{align}
Finally, the screened defect potential in real space can be calculated
\begin{equation}
    \label{eqn:pot}
    V_{\text{mono}}^{\pm}(r) =
    \frac{1}{2\pi}\int\mathrm{d}q\; V^{\pm}_{\text{mono}}(q)J_0(qr),
\end{equation}
where $J_0$ is the zeroth Bessel function of the second kind.

To evaluate the screened defect potential, we determine $\chi_0(q)$ using the first-principles random-phase approximation (RPA). For this, we first calculate Kohn-Sham wavefunctions $\left\lbrace|\phi_{n\mathbf{k}}\rangle\right\rbrace$ and energies $\left\lbrace E_{n\mathbf{k}}\right\rbrace$ of all occupied states and 1490 unoccupied states on a $13\times13$ $k$-point mesh for a defect-free BP monolayer using plane-wave pseudopotential density-functional theory (DFT) as implemented in the \textsc{Quantum Espresso (v6.5)} software package \cite{Giannozzi2009}. For these calculations, the PBE (GGA) exchange-correlation functional \cite{Perdew1996}, optimized norm-conserving Vanderbilt (ONCV) pseudopotentials \cite{Hamann2013} and a 120~Ry plane-wave cutoff were used. We then determine the inverse dielectric matrix $\varepsilon_{\mathbf{G}\mathbf{G'}}^{-1}(\mathbf{q})$ as a function of $\mathbf{q}$ in the first Brillouin zone (BZ), where $\mathbf{G}$ and $\mathbf{G'}$ are reciprocal lattice vectors. This is achieved using the \textsc{BerkeleyGW} package \cite{deslippe2012}, by evaluating the Adler-Wiser formula \cite{adler, wiser, Qiu2016}:
\begin{equation}
    \varepsilon_{\mathbf{G}\mathbf{G'}}^{-1}(\mathbf{q}) = \delta_{\mathbf{G}\mathbf{G'}} + \frac{v_{\text{tr}}(\mathbf{q}+\mathbf{G'})}{\Omega}\sum_{cv\mathbf{k}}\frac{M_{vc\mathbf{k}}(\mathbf{q},\mathbf{G})M_{cv\mathbf{k}}^*(\mathbf{q},\mathbf{G'})}{E_{c\mathbf{k+q}} - E_{v\mathbf{k}}},
\end{equation}
where $M_{nn'\mathbf{k}}^*(\mathbf{q},\mathbf{G})=\langle\phi_{n\mathbf{k+q}}|e^{i(\mathbf{q}+\mathbf{G})\cdot\mathbf{r}}|\phi_{n'\mathbf{k}}\rangle$, $\Omega$ is the crystal volume (given by the unit cell volume times the number of $k$-points) and  $v_{\text{tr}}(\mathbf{q})$ is the 3D Fourier transform of the slab-truncated Coulomb interaction~ \cite{Ismail-Beigi2006}. For the RPA calculation, a plane-wave cutoff of 15 Ry was used. 

The anisotropic 2D inverse dielectric function and the anisotropic non-interacting response function $\chi^\text{ani}_0(\mathbf{q})$ are then given by~\cite{Qiu2016}
\begin{align}
    \varepsilon^{-1}_{\text{2D}}(\mathbf{q}) & =\frac{1}{1 - v_q\chi^\text{ani}_0(\mathbf{q})} \\
    & = \frac{q}{2\pi e^2 L_z}\sum_{\mathbf{G}_z\mathbf{G}^{\prime}_z}\varepsilon^{-1}_{\mathbf{G}_z\mathbf{G}^{\prime}_z}(\mathbf{q})v(\mathbf{q}+\mathbf{G}^{\prime}_z),
\label{eq:epseff}
\end{align}
where $L_z=18$~\AA\ is the distance between periodically repeated BP monolayers. We find that $\chi^\text{ani}_0(\mathbf{q})$ only varies by less than three percent as function of the azimuthal angle justifying our assumption of isotropic screening above. Finally, the isotropic effective response function $\chi_0(q)$, which is used to calculate the screened defect potential in Eq.~\eqref{eq:Vind}, is obtained as the angular average of the anisotropic response function.

\subsection{Bilayer BP}

Next, we discuss the calculation of the screened defect potential in the BP bilayer. Again, we express this potential as the sum of an unscreened potential and an induced potential. The induced potential is obtained from the induced densities $\delta n_{\text{ind}}^{(l)}(q)$ in the two monolayers, with $l=1,2$ denoting the monolayer. Again, we assume that these induced densities are located in the effective sheets located between the sublayers of each monolayer, indicated by the red dashed lines in the inset of Fig.~\ref{fig:mono_pot} (b). The induced densities can be expressed in terms of the non-interacting monolayer response functions $\tilde{\chi}_0(q)$ (which is different from $\chi_0(q)$ because the monolayers are part of a bilayer) via
\begin{align}
    \delta n_{\text{ind}}^{(1)}(q) &= \frac{(\tilde{\chi}^{-1}_0(q)-v_q)V_0^{(1)}(q) + v_qe^{-qh}V_0^{(2)}(q)}{(\tilde{\chi}^{-1}_0(q)-v_q)^2 - v^2_qe^{-2qh}},\\
    \delta n_{\text{ind}}^{(2)}(q) &= \frac{(\tilde{\chi}^{-1}_0(q)-v_q)V_0^{(2)}(q) + v_qe^{-qh}V_0^{(1)}(q)}{(\tilde{\chi}^{-1}_0(q)-v_q)^2 - v^2_qe^{-2qh}},
\end{align}
where $h=5.24$~\AA\ is the distance between the two effective sheets and $V_0^{(l)}$ is the unscreened defect potential experienced by the effective sheets. 

The total potential is then given by
\begin{equation}
    V_{\text{bi}}^{(l),\pm}(q) = V^{(l),\pm}_{0}(q) + \sum_{l'=1}^2 v_{q}e^{-q|h^{\pm}_{l} - z_{l'}|}\delta n^{(l')}_{\text{ind}}(q),
\label{eq:Vbi}
\end{equation}
where $V^{(l),\pm}_{0}(q)$ is the unscreened defect potential (with $\pm$ denoting the two sublayers of monolayer $l$), $z_l$ denotes the $z$-coordinate of the effective sheet associated with monolayer $l$ and $h^{\pm}_{l}$ is the $z$-coordinate of the top ($+$) or bottom ($-$) sublayer of monolayer $l$.

To determine $\tilde{\chi}_0(q)$, we compare the screened potential between two charges located in the same effective sheet from our screening model, where $V_0^{(l)}=v_qe^{-qh(l-1)}$, to the result of a first-principles calculation. According to Eq.~\eqref{eq:Vbi}, this potential is given by 
\begin{equation}
    \label{eq:bilayer_model}
    V_{\text{scr}}(q) = v_q\left[\frac{\tilde{\chi}_0^{-2}(q) - \tilde{\chi}_0^{-1}(q)v_q(1 - e^{-2qh})}{\tilde{\chi}_0^{-2}(q) - 2\tilde{\chi}_0^{-1}(q)v_q + v_q^2(1 - e^{-2qh})}\right].
\end{equation}

To calculate this potential from first principles, we follow the same steps as in the monolayer case \footnote{For the DFT calculations, ONCV pseudopotentials for phosphorus are used, but we instead use a plane-wave cutoff of 90~Ry, an out-of-plane periodicity of 20~\AA, and a $13\times13$ $k$-grid. Then for Eq. \ref{eq:epseff} we use a cutoff of 11~Ry and include approximately 1200 states.} to determine  $\varepsilon^{-1}_{\text{2D}}(q)$ for the bilayer. Solving $\varepsilon^{-1}_{\text{2D}}(q)v_q=V_{\text{scr}}(q)$ for $\tilde{\chi}_0^{-1}(q)$ then yields 
\begin{align}
    v_q^{-1}\tilde{\chi}_0^{-1}(q) &= \frac{\varepsilon_{\text{2D}}(q)(1 - e^{-2qh}) - 2 - \mathcal{M}(q, h) }{2(\varepsilon_{\text{2D}}(q) - 1)},\\
    \mathcal{M}(q, h) & = \sqrt{\varepsilon_{\text{2D}}^2(q)(1 - e^{-2qh})^2 + 4e^{-2qh}}.
\end{align}

Figures~\ref{fig:mono_pot}(a) and (b) show the sublayer-resolved screened defect potential for a charged adsorbate on monolayer and bilayer BP. As expected for 2D semiconductors, the screened potential in all sublayers approaches the unscreened potential at large distances from the adsorbate. However, at short distances the screened potential is reduced significantly. In bilayer BP, the screened potential in the top monolayer behaves qualitatively similarly to the monolayer case, but is somewhat weaker as a consequence of screening from the additional BP layer. In the bottom monolayer the screened potential is an order of magnitude smaller than in the top layer and relatively flat indicating that bound defect states should reside in the top monolayer. 

In addition to adsorbates, we also model charged defects that are intercalated between the two monolayers of bilayer BP. In particular, we consider a charge placed equidistant from two P atoms in the top layer and two P atoms in the bottom layer, see Fig.~\ref{fig:structure}(b), which is a typical configuration of intercalated transition metal atoms, such as copper or nickel~\cite{Gaberle2018}.

\begin{table}[b!]
    \centering
    \begin{tabular}{c||c|c||c|c}
        & \multicolumn{2}{c||}{Monolayer} & \multicolumn{2}{c}{Bilayer} \\
         &  Electron & Hole & Electron & Hole\\
        \hline
        $m^*_{\text{a}}/m_e$ &  0.196  & 0.166 &  0.180 & 0.152 \\
        $m^*_{\text{z}}/m_e$ &  1.160 &  3.241 &  1.530 &  1.362 \\
    \end{tabular}
    \caption{Effective masses along the armchair ($m_{\text{a}}$) and zigzag ($m_{\text{z}}$) directions at the $\Gamma$ point for monolayer and bilayer BP from the tight-binding model of Rudenko and coworkers~\cite{Rudenko2015} calculated using a fourth-order finite difference approximation to the second derivative of the bands. Note that $m_e$ denotes the free electron mass.}
    \label{tab:eff_mass}
\end{table}

\begin{figure}
    \centering
    \includegraphics[width=0.495\textwidth]{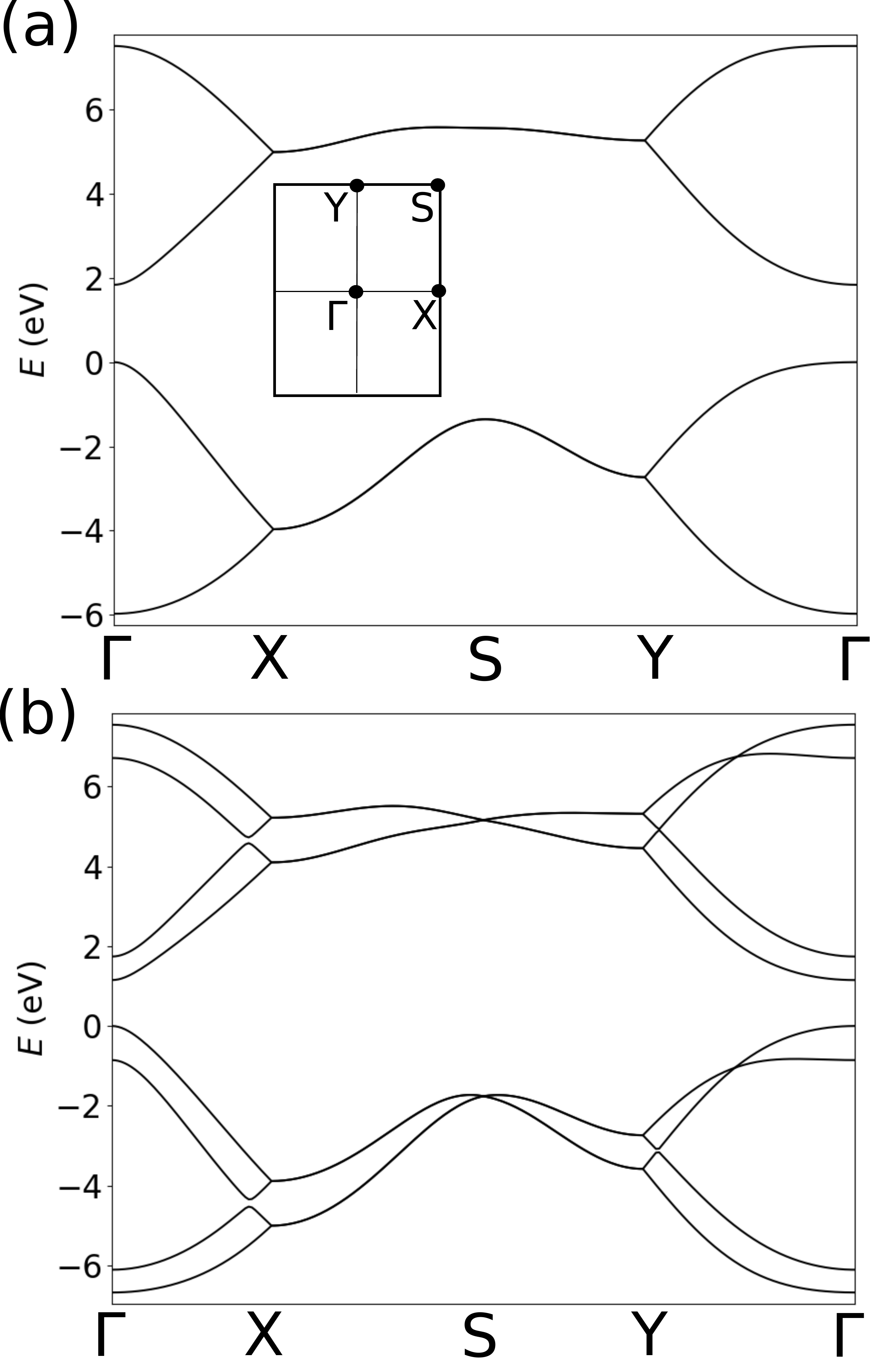}
    \caption{Electronic band structure of (a) monolayer and (b) bilayer black phosphorus from tight-binding. The inset of (a) indicates the high symmetry points of the Brillouin zone. All energies are referenced to the valence band maximum.}
    \label{fig:bands}
\end{figure}

To study the behaviour of electrons in the presence of a charged defect, we use the tight-binding model of Rudenko \emph{et al.}~\cite{Rudenko2015} which includes hoppings between phosphorus $p_z$ orbitals. The band structures of monolayer and bilayer BP from this approach are shown in Fig.~\ref{fig:bands}. For the monolayer a direct band gap at $\Gamma$ of 1.84~eV is obtained, which is consistent with $GW_0$ calculations~\cite{Rudenko2015}. For the bilayer, the band gap is reduced to 1.15~eV as a consequence of interlayer hopping. Importantly, both the monolayer and the bilayer exhibit highly anisotropic effective masses in both the conduction and valence band. Table~\ref{tab:eff_mass} shows that the effective masses in the zigzag direction $m_{\text{z}}$ are about one order of magnitude larger than in the armchair direction $m_{\text{a}}$.

Next, the screened defect potential is included as an additional on-site potential in the tight-binding Hamiltonian. To describe the slow decay of the screened defect potential, a $60\times60$ supercell is used containing 14,400 P atoms for the monolayer and 28,800 P atoms in the bilayer calculations. We sample at the $\Gamma$ point to obtain the binding energies and wavefunctions of the defect supercell, and use a $3\times3$ $k$-point sampling to obtain the local density of states (LDOS).

We compare the results of our atomistic calculations to a variational calculation based on the effective mass approximation. In this method, the following ansatz is made for the most strongly bound defect state
\begin{equation}
    \Psi_{1s}(\mathbf{r}; \alpha, \beta) = \sqrt{\frac{2\alpha\beta}{\pi}}e^{-\sqrt{(\alpha x)^2 + (\beta y)^2}},
    \label{eq:EMA}
\end{equation}
where $\mathbf{r}=(x,y)$ and $\alpha,\,\beta$ are parameters that describe the decay of the defect state. Note that this form of the wavefunction does not describe the sublayer structure of the BP monolayer.

To enable a quantitative comparison between the binding energies of defect states obtained from the atomistic calculations and the effective mass approach, the position of the plane in which the wavefunction in Eq.~\eqref{eq:EMA} is localized must be chosen. As Fig.~\ref{fig:mono_pot} shows that the screened defect potential has a significantly larger value in the top sublayer than in the other sublayers, we assume that the wavefunction of the most strongly bound $1s$ state is primarily localized in the top sublayer. 

The expectation value of the effective mass Hamiltonian is then given by 
\begin{align}
\label{eqn:var2}
    & \langle\Psi_{1s}(\alpha, \beta)|H|\Psi_{1s}(\alpha, \beta)\rangle  = \frac{\hbar^2}{4}\left(\frac{\alpha^2}{m^{*}_{\text{a}}} + \frac{\beta^2}{m^{*}_{\text{{z}}}}\right) + V(\alpha,\beta), \\
    & V(\alpha,\beta)  = \frac{2}{\pi}\int\mathrm{d}r\; re^{-2r}\int\mathrm{d}\theta\;  V_{\text{def}}\left(\left|\left(\frac{r\text{cos}\theta}{\alpha}, \frac{r\text{sin}\theta}{\beta}\right)\right|\right),\\
    & V_{\text{def}}(r) = 
    \frac{e^2}{4\pi\varepsilon_0}\int\mathrm{d}q\;\frac{J_0(qr)e^{-qd}}{\frac{\varepsilon_{\text{sub}} - 1}{2} + \varepsilon_{\text{2D}}(q)},
\end{align}
where $V_{\text{def}}$ is the defect potential.

\section{\label{sec:part1}Monolayer Black Phosphorus}

\subsection{\label{subsec:wfc1}Bound State Wavefunctions}

\begin{figure*}[ht!]
    \centering
    \includegraphics[width=1.0\textwidth]{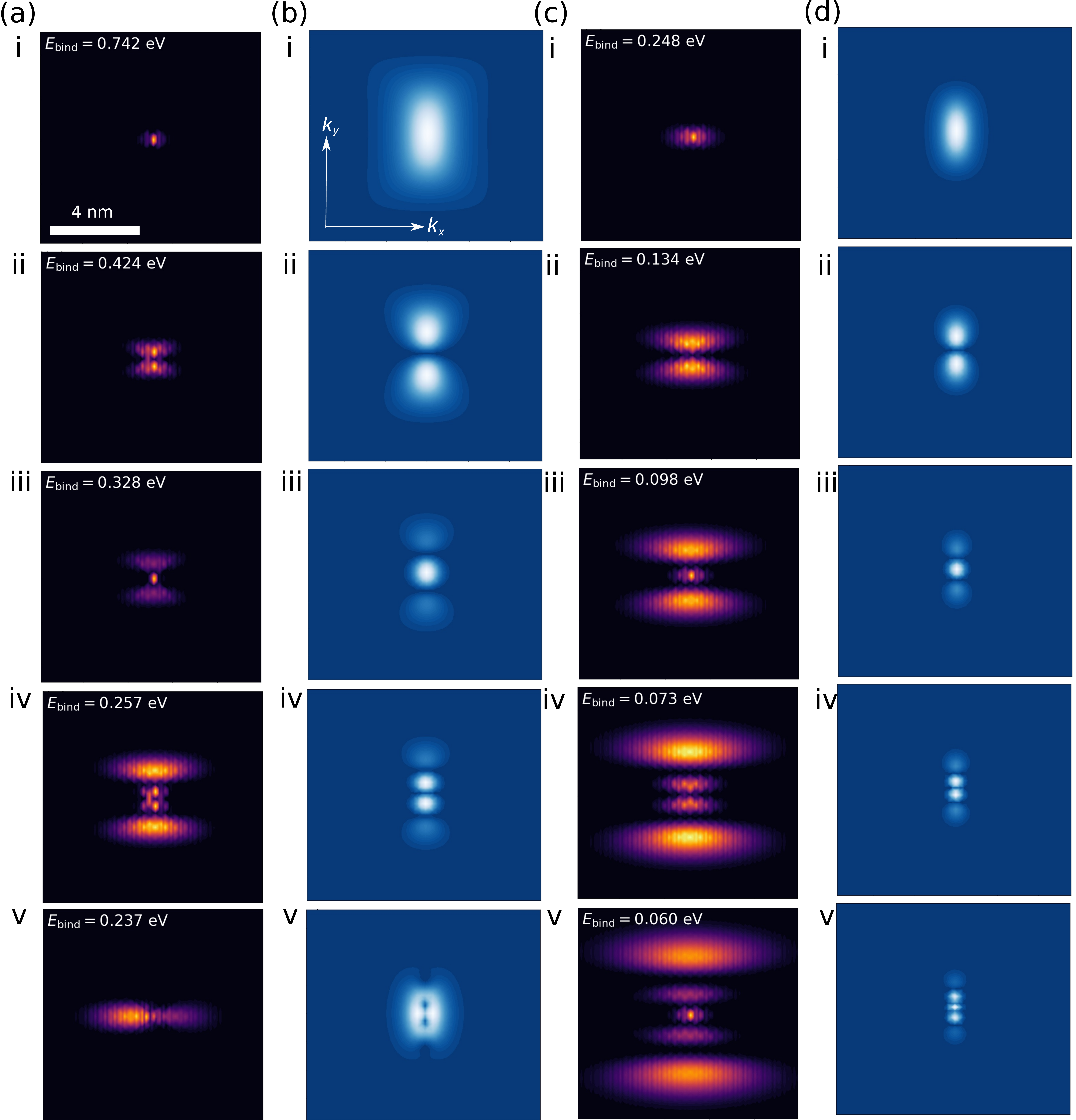}
    \caption{(a, c)(i-v) Wavefunctions of the five most strongly bound acceptor states in monolayer black phosphorus and (b, d)(i-v) their projections onto pristine monolayer states in the first Brillouin zone, with the $\Gamma$-point located in the centre of each panel. The substrate dielectric constant is $\varepsilon_{\text{sub}}=2$ in columns (a) and (b) and $\varepsilon_{\text{sub}}=8$ in columns (c) and (d). The adsorbate is a single negative charge at a height of $d=2$~\AA\ in all cases. All defect binding energies $E_{\text{bind}}$ are given relative to the valence band edge.}
    \label{fig:wfn1}
\end{figure*}

Figure~\ref{fig:wfn1} shows the wavefunctions of the bound defect states in monolayer BP induced by a negatively charged acceptor adsorbate with a height of $d=2$~\AA\ above the top sublayer. Figs.~\ref{fig:wfn1}(a)(i-v) show results for a substrate with $\varepsilon_{\text{sub}}=2$, while Fig.~\ref{fig:wfn1}(c)(i-v) shows results for $\varepsilon_{\text{sub}}=8$. As a consequence of the large difference of the effective masses in the armchair and zigzag directions, these wavefunctions resemble anisotropically distorted 2D hydrogenic orbitals and we therefore label them as $1s$ (i), $2p_y$ (ii), $2s$ (iii), $3p_y$ (iv), and $2p_x$ (v). Note that the ordering of the defect states is different than in the 2D hydrogen atom as a consequence of the anisotropic dispersion which lowers the energy of states that are elongated along the zigzag direction compared to those along the armchair direction. For example, the $2p_x$ and the $2p_y$ orbitals are degenerate in the 2D hydrogen atom, but in monolayer BP the binding energy of the $2p_y$ state is almost 0.2~eV larger than that of the $2p_x$ state. Projecting the defect wavefunctions onto the unperturbed monolayer states in the first Brillouin zone, as shown in Figs.~\ref{fig:wfn1}(b)(i-v) for the states shown in Figs.~\ref{fig:wfn1}(a)(i-v) and in Figs.~\ref{fig:wfn1}(d)(i-v) for the states shown in Figs.~\ref{fig:wfn1}(c)(i-v), reveals that they are predominantly constructed from bulk valence states near the valence band maximum at the $\Gamma$-point. Importantly, we find that the projections are delocalized over a large part of the first Brillouin zone.

Figure~\ref{fig:wfn1} also demonstrates that increasing the substrate dielectric constant from $\varepsilon_{\text{sub}}=2$ to $\varepsilon_{\text{sub}}=8$ causes the defect states to become more delocalized and their projections to become more localized to the vicinity of the $\Gamma$-point. We also observe changes in the ordering of the defect states with the fifth state exhibiting a different symmetry for $\varepsilon_{\text{sub}}=8$ as compared to the case of $\varepsilon_{\text{sub}}=2$.

Results for a positively charged donor adsorbate are shown in the appendix. The donor states are similar in shape and spatial extent to the acceptor states, but have a different ordering of the $2p_x$ state relative to the $3p_y$ state.

\begin{figure*}[ht!]
    \centering
    \includegraphics[width=0.95\textwidth]{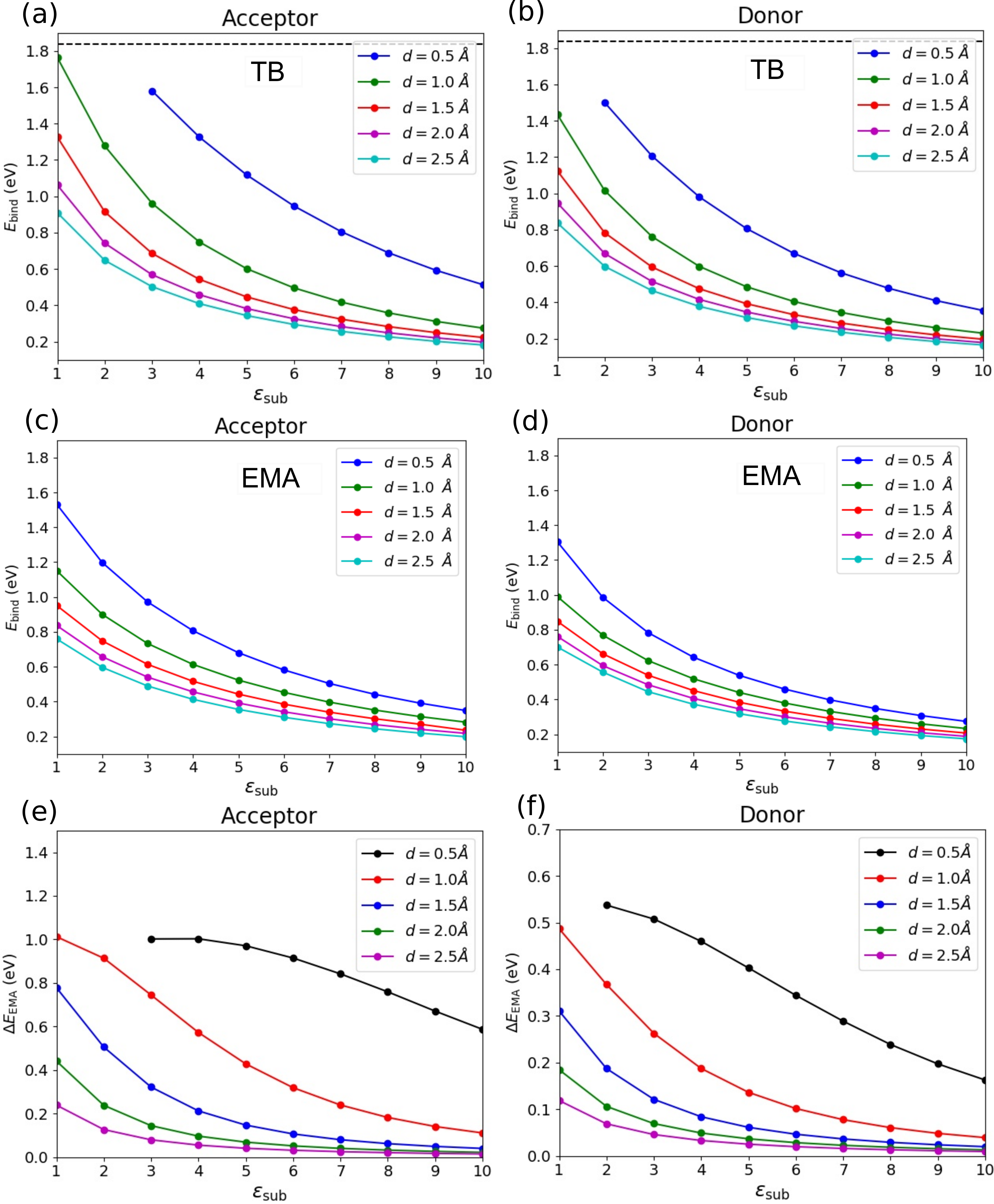}
    \caption{Binding energy $E_{\text{bind}}$ of (a) donor and (b) acceptor $1s$ defect states as a function of substrate dielectric constant $\varepsilon_{\text{sub}}$ and defect height $d$, obtained from atomistic tight-binding (TB) calculations. The band gap $E_{\text{g}}$ is shown as a black dashed line, and the data points for the $1s$ state binding energies that exceed $E_{\text{g}}$ have been omitted. (c) and (d): Corresponding results from the variational effective mass approximation (EMA). Average parabolic energy deviation $\Delta E_{\text{EMA}}$ of the $1s$ state (see Eq.~\ref{eq:deviation}) for the case of (e) acceptor and (f) donor, shown as a function of $\varepsilon_{\text{sub}}$ for the same $d$ values as in (a-d).}
    \label{fig:energy_var}
\end{figure*}
\subsection{\label{subsec:energies1}Binding Energies}

\begin{figure*}[ht!]
    \centering
    \includegraphics[width=0.95\textwidth]{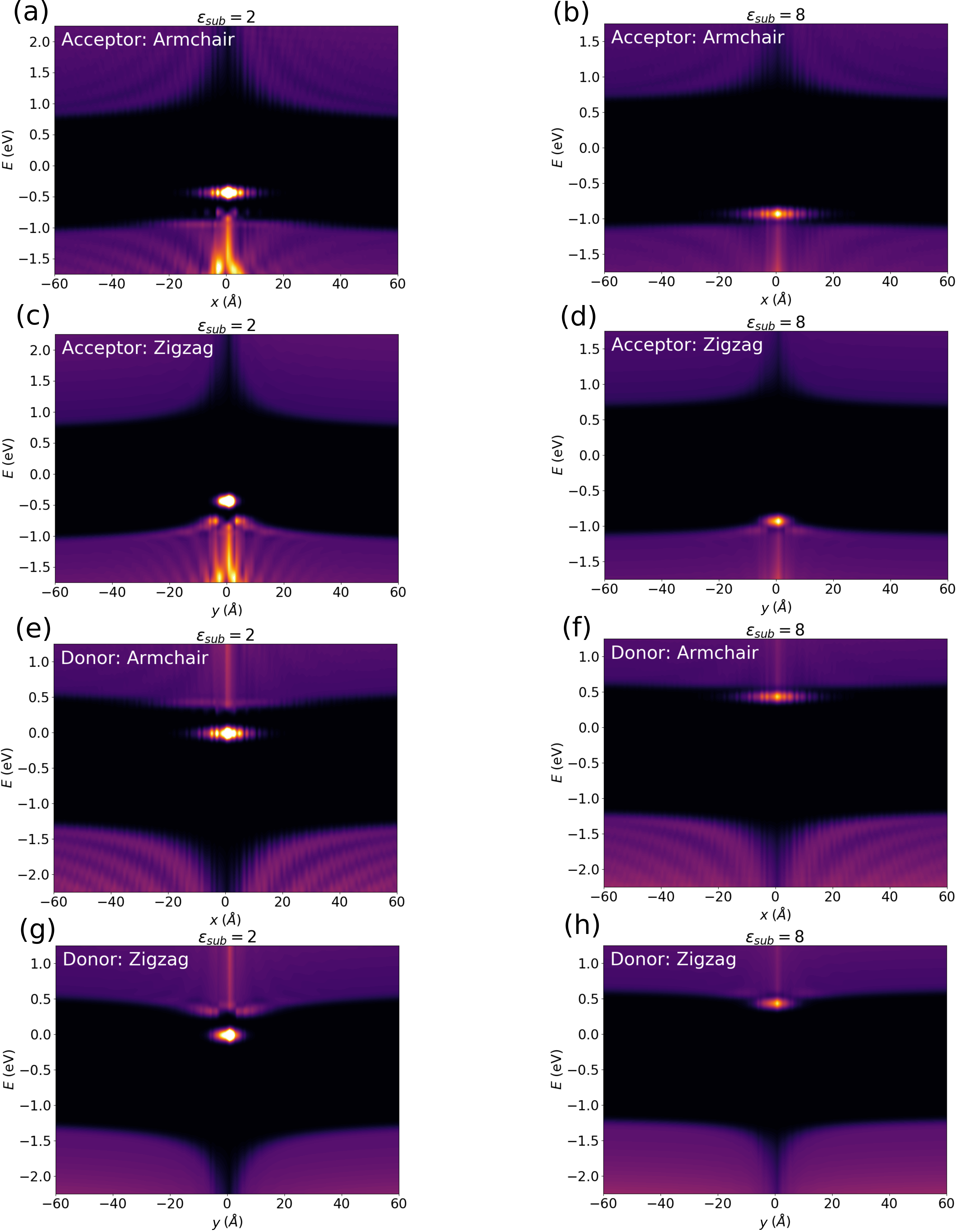}
    \caption{Local density of states for charged adsorbates (with $d=2$~\AA) on monolayer BP with (left column) $\varepsilon_{\text{sub}}=2$ and (right column) $\varepsilon_{\text{sub}}=8$, for (a-d) acceptor defects and (e-h) donor defects. Results are shown both along the armchair and and zigzag directions.}
    \label{fig:ldos}
\end{figure*}

Figure~\ref{fig:energy_var} shows the binding energies of the $1s$ defect state as a  function of the defect height $d$ and the substrate dielectric constant $\varepsilon_{\text{sub}}$. As expected, an increase in $\varepsilon_{\text{sub}}$ reduces the binding energies as the defect potential is weaker. Interestingly, we find for small defect heights ($d=1.5$~\AA) and weakly screening substrates ($\varepsilon_{\text{sub}}<3$) that the binding energy of the acceptor state exceeds the size of the band gap, see Fig.~\ref{fig:energy_var}(a). In this case, the defect state becomes resonant with the conduction band states. Similarly, the donor states can become resonant with valence band states. We find that binding energies of acceptors are generally larger than those of donor impurities. This can be attributed to the larger value of $m^{*}_{\text{z}}$ at the valence band maximum compared to the conduction band minimum, see Table~\ref{tab:eff_mass}.

Figures~\ref{fig:energy_var}(c-d) show the binding energies obtained from the variational effective mass approach, see Eq.~\eqref{eq:EMA}. We find that the results from this approach are generally in good qualitative agreement with the atomistic calculations, but there are some important differences. In particular, we find that the effective mass approximation underestimates binding energies when the defect is close to the monolayer and the substrate has a small dielectric constant. Most importantly, the effective mass approximation does not reproduce the result that the most strongly bound defect state can cross the entire band gap and become resonant with bulk states. 

This failure of the effective mass approximation can be understood by analyzing the projection of the defect state onto bulk monolayer states. As discussed above (see Fig.~\ref{fig:wfn1}), the projections for the most strongly bound $1s$ state are delocalized over a larger part of the first Brillouin zone. However, the effective mass approximation is based on the assumption of an anisotropic \emph{parabolic} dispersion which is only valid near the conduction and valence band edges. 

To quantitatively assess the validity of the effective mass approximation, we calculate  
\begin{equation}
\label{eq:deviation}
    \Delta E_\text{EMA}  \equiv \sum_{\mathbf{k}} \left| \langle n\mathbf{k} | \Psi_{1s} \rangle \right|^2\left|E_{n \mathbf{k}} - \frac{\hbar^2k_x^2}{2m^{*}_{\text{a}}} - \frac{\hbar^2k_y^2}{2m^{*}_{\text{z}}}\right|,
\end{equation}
which measures the importance of deviations from a parabolic dispersion for the defect state binding energies. In this equation, $\langle n\mathbf{k} | \Psi_{1s} \rangle$ denotes the projection of the most strongly bound $1s$ defect state (from the atomistic calculation) onto the pristine monolayer state $|n\mathbf{k}\rangle$ (where the band index $n$ refers to the highest valence band for acceptor states and the lowest conduction band for donor states). Also, $\mathbf{k}$ denotes a crystal momentum and $E_{n\mathbf{k}}$ is the tight-binding band structure of the monolayer.

Figures~\ref{fig:energy_var}(e) and (f) show $\Delta E_\text{EMA}$ for the most strongly bound acceptor and donor states. As expected, we find that $\Delta E_\text{EMA}$ is large for small defect heights and small values of the substrate dielectric constant. Comparing donor and acceptor states, we observe that $\Delta E_\text{EMA}$ is larger for acceptor states. This is because the lowest conduction bands are more parabolic and can therefore be accurately described by an anisotropic effective mass model over a larger region of the Brillouin zone than the valence states.

\subsection{Local Density of States}

Figure~\ref{fig:ldos} shows the local density of states (LDOS) for monolayer BP with positively or negatively charged adsorbates along the zigzag and armchair directions. The LDOS is directly accessible in scanning tunneling spectroscopy (STS) measurements. For a substrate with $\varepsilon_{\text{sub}}=2$, an acceptor defect gives rise to a strong peak several tenths of an eV above the valence band maximum which originates from the $1s$ bound state, see Figs.~\ref{fig:ldos}(a) and (c). This peak is followed by a series of closely spaced peaks near the valence band maximum. Strong defect-induced band bending can be observed in the conduction band. Comparing the acceptor LDOS along the armchair and zigzag directions, we find that peaks are more extended along the armchair directions reflecting the anisotropy of the defect wavefunctions, see Fig.~\ref{fig:wfn1}. For a substrate with $\varepsilon_{\text{sub}}=8$, the $1s$ peak is much closer to the valence band and the additional peaks from less strongly bound defect states cannot be observed clearly, see Figs.~\ref{fig:ldos}(b) and (d). For donor adsorbates, we observe peaks near the conduction band arising from bound states as well as a downward bending of the valence band.

\section{\label{sec:part2}Bilayer Black Phosphorus}
\subsection{\label{subsec:wfc2}Bound State Wavefunctions}

\begin{figure}[ht!]
    \centering
    \includegraphics[width=0.475\textwidth]{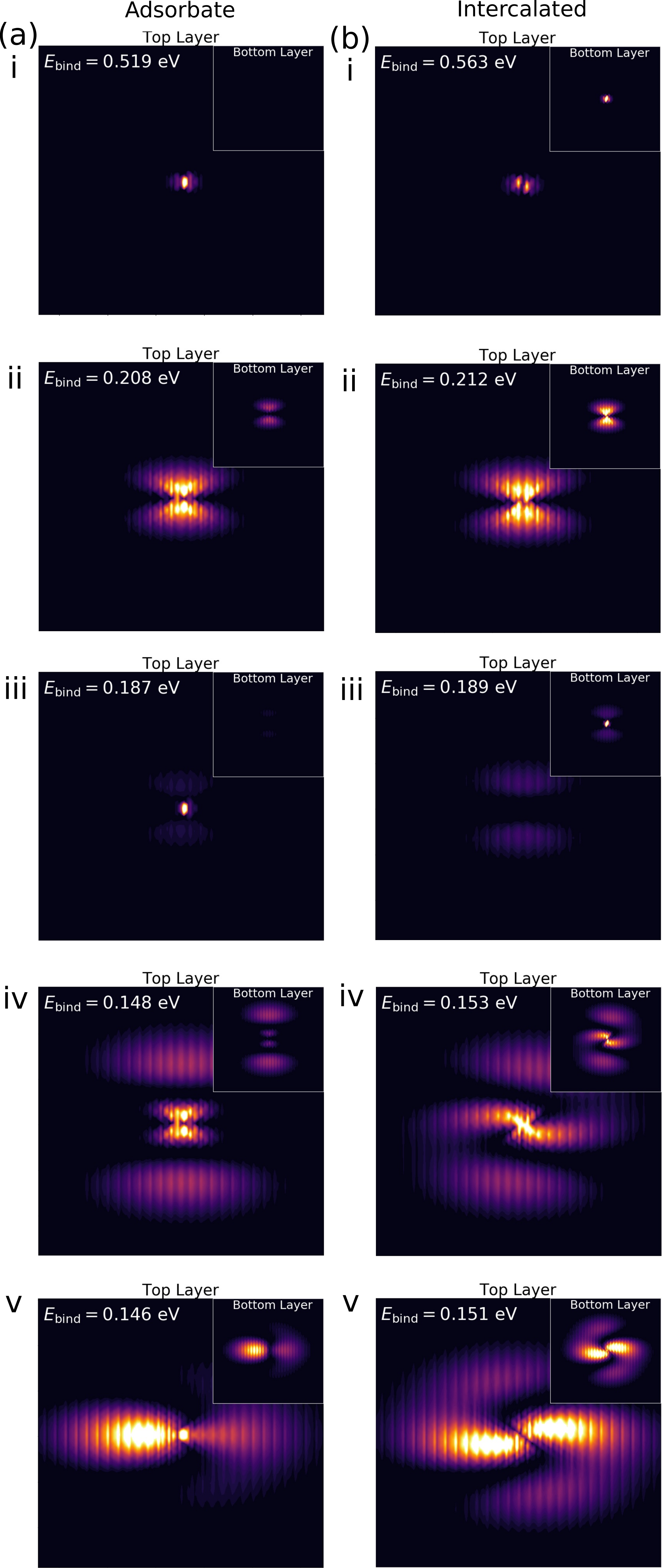}
    \caption{Wavefunctions of the five most strongly bound acceptor states in bilayer BP in the top monolayer. (a) Results for a negatively charged adsorbate $d=2$~\AA\ above the top sublayer and $\epsilon_{\text{sub}}=1$ and (b) results for a negatively charged intercalant. The wavefunctions in the bottom monolayer are shown in the insets.}
    \label{fig:bilayer_wfc}
\end{figure}

Figure~\ref{fig:bilayer_wfc}(a) shows the most strongly bound defect states of a negatively charged adsorbate on bilayer BP. The defect wavefunctions have similar shapes and nodal structures as in the monolayer case, but a larger spatial extent. The wavefunctions in the bottom layer have a similar shape as in the top layer but the intensity is significantly reduced as a result of the screening by the top layer, see Fig.~\ref{fig:mono_pot}(b). Like in the monolayer, the bilayer defect states are primarily composed of $\Gamma$-point states of the pristine bilayer. We find similar results for the case of a donor adsorbate, but the order of states iv and v in Fig.~\ref{fig:bilayer_wfc}(a) is switched, see appendix. 

The defect wavefunctions of a negatively charged intercalant are shown in Fig.~\ref{fig:bilayer_wfc}(b). In contrast to the adsorbate case, the wavefunctions for this system have a similar intensity on both layers. Interestingly, the wavefunction of the most strongly bound state exhibits two peaks in the top layer, but only a single peak in the bottom layer, see Fig.~\ref{fig:bilayer_wfc}(b)i. The splitting of this state into two peaks is an interference effect caused by the location of the intercalant below the centre of a zigzag bond, see Fig.~\ref{fig:structure}. The three most strongly bound states of the acceptor intercalant have a similar nodal structure as in the adsorbate case. However, the fourth and the fifth states look very different and appear to be mixtures of the corresponding adsorbate states (which are very close in energy).

\subsection{\label{subsec:energies2}Binding Energies}

\begin{figure}
    \centering
    \includegraphics[width=0.39\textwidth]{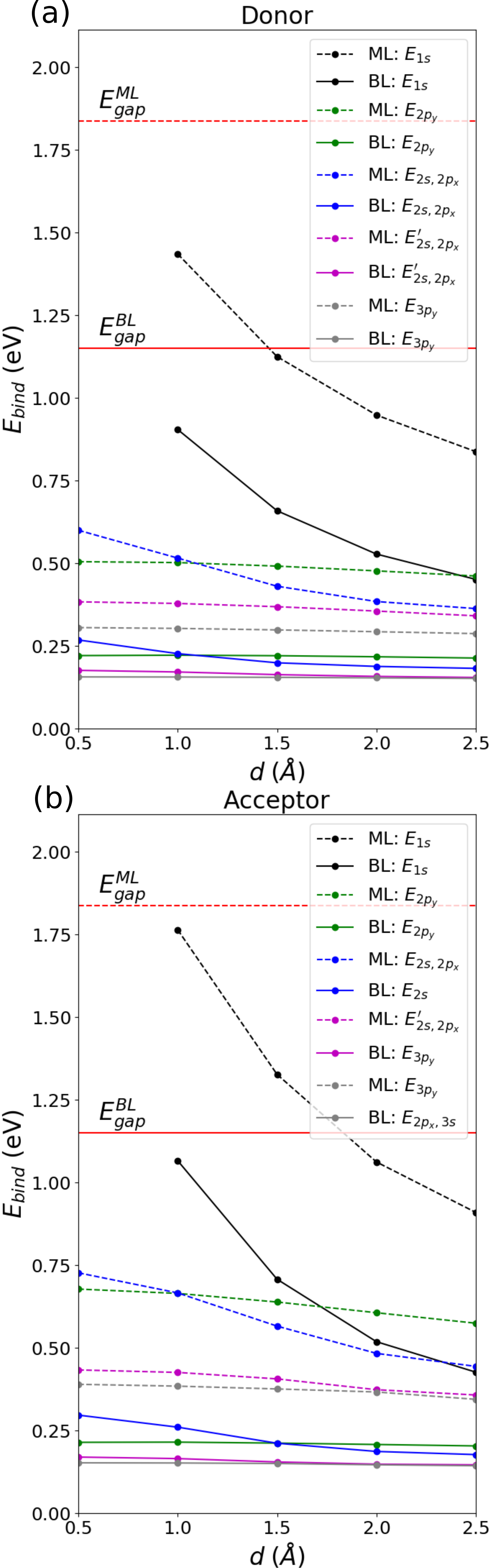}
    \caption{Comparison of binding energies of various bound defect states in freestanding monolayer BP (dashed lines) and freestanding bilayer BP (solid lines) for (a) donor and (b) acceptor adsorbates. The band gap of the monolayer and bilayer are shown as red dashed and red solid horizontal lines, respectively. Note that $E_{2s,2p_{x}}$ denotes the binding energy of a bound state that is a mixture of $2s$ and $2p_x$-like states. If two states with a similar character are found, one of them is indicated by a prime.}
    \label{fig:bilayer_energies}
\end{figure}

\begin{figure*}
    \centering
    \includegraphics[width=0.99\textwidth]{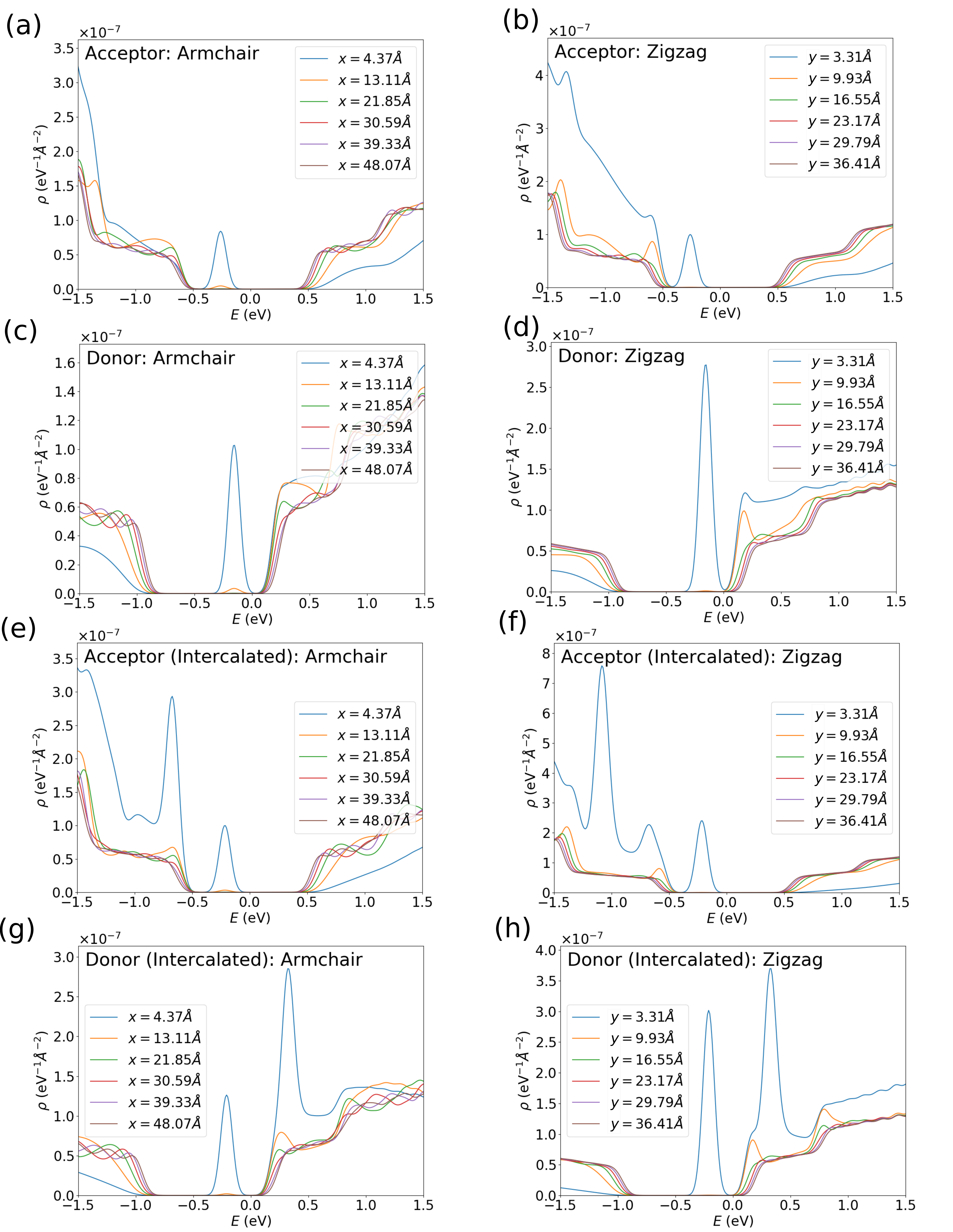}
    \caption{(a)-(d): Local density of states (LDOS) of a charged adsorbate placed 2~\AA\ above the top sublayer of the top layer of bilayer BP, shown for the acceptor case along the (a) armchair and (b) zigzag directions, and shown for the donor case along the (c) armchair and (d) zigzag directions. (e)-(h) Local density of states (LDOS) of a charged intercalant, shown for the acceptor case along the (e) armchair and (f) zigzag directions, and shown for the donor case along the (g) armchair and (h) zigzag directions.}
    \label{fig:bl_ldos}
\end{figure*}

Figure~\ref{fig:bilayer_energies} shows the binding energies of the most strongly bound defect states of both acceptor and donor adsorbates on freestanding ($\varepsilon_{\text{sub}}=1$) bilayer BP as function of the defect height. For comparison, results for the freestanding monolayer are also shown. As expected, the binding energies decrease with increasing defect height but the decrease is most pronounced for the $1s$ state. For very small defect heights, the binding energy of the $1s$ state exceeds the band gap and the state becomes resonant with the valence band (in case of donor impurities) or the conduction  band (in case of acceptors). The behaviour of the $1s$ binding energy is a consequence of its localized wavefunction which is strongly confined to the direct vicinity of the defect. As a consequence, the electrons in the $1s$ state only experience the defect potential close to the defect which decays as $1/d$. Comparing the binding energies of the monolayer and the bilayer case, we find that the monolayer binding energies of the $1s$ state are $\sim0.4-0.5$~eV larger than in the bilayer case and the difference is insensitive to the defect height. 

In contrast, the binding energies of the other states (with only the exception of the $2s$ state which is also strongly localized) are less sensitive to the defect height as their wavefunctions are much more delocalized. Again, the monolayer states behave similarly but their binding energies are about 0.5~eV larger than for the bilayer.

\section{Local Density of States}

Figures~\ref{fig:bl_ldos}(a) and (b) show the LDOS $\rho(E, \mathbf{r})$ of an acceptor adsorbate on bilayer BP as a function of energy for several distances from the defect along the armchair and zigzag directions, respectively. The LDOS exhibits a prominent peak in the gap which can be attributed to the strongly bound and localized $1s$ state. As a result of BP's anisotropy, the peak decays less rapidly along the armchair direction. Additional peaks are found close to the valence band edge which can be clearly seen along the zigzag direction. The acceptor defect induces a flattening of the LDOS associated with conduction band states arising from band bending~\cite{Aghajanian2018, Aghajanian2020, Corsetti2017}. As the distance from the defect increases, the LDOS converges to that of the pristine bilayer. 

For the donor case, shown in Fig.~\ref{fig:bl_ldos}(c) and (d) along armchair and zigzag directions, respectively, we again find a prominent peak arising from the $1s$ state as well as smaller additional features at the conduction band edge. 

The LDOS of charged intercalants are shown in Figs.~\ref{fig:bl_ldos}(e-h). Again, the most strongly bound state gives rise to a prominent in-gap peak, but in contrast to the adsorbate case, we see stronger additional peaks along both the zizag and armchair directions. These additional peaks are resonant with the valence (conduction) band for acceptor (donor) intercalants and decay quickly as the distance from the defect is increased indicating that they originate from localized defect states.  

The different electronic properties of charged adsorbates and charged intercalants in bilayer BP can be understood from an analysis of the band structure, see Fig.~\ref{fig:bands}(b). The difference in electronic response between a defect adsorbed to the top layer and one intercalated between the layers can be explained in terms of the secondary conduction and valence band extrema at the $\Gamma$-point in the bilayer system, seen in Fig.~\ref{fig:bands}(b). In particular, the hybridization between the highest valence (conduction) bands of the two monolayers gives rise to a pair of valence (conduction) bands near $\Gamma$ with similar dispersion but an energy splitting of 0.85~eV for the valence bands and 0.59~eV for the conduction bands. When a charged defect is present, each of these bands forms a set of distorted hydrogenic defect states~\cite{Aghajanian2018,Bassani_1974}, but the defect states arising from the secondary valence (conduction) bands are resonant with the delocalized valence band states. Moreover, the secondary valence and conduction band states are  moderately more composed from the orbitals of the inner P atoms which the intercalant is closer to, while the highest lying valence band and the lowest lying conduction band are more composed from the orbitals of the outer P atoms. As a consequence, charged intercalants interact more strongly with those secondary valence and conduction band states than in the adsorbate case, which explains the stronger secondary peaks observed in the LDOS.

\section{Conclusions}
In this paper, we have used large-scale tight-binding simulations to study the electronic structure of monolayer and bilayer black phosphorus (BP) with charged defects. The screened potential induced by the defect is obtained from first-principles linear-response theory and decays slowly necessitating the use of large supercells containing thousands of atoms. The presence of the defect potential gives rise to the formation of bound states that resemble anisotropically distorted hydrogenic orbitals. We have studied the binding energies of these states as a function of the defect position and also of the dielectric constant of the substrate. In the monolayer, the binding energy of acceptor states is larger than that of donor states as a consequence of the different effective masses in the conduction and valence bands. We compare the results of our atomistic calculations to an effective mass model and find significant differences for small defect heights. When the defect is close to the BP layer, its wavefunction contains contributions from states in a large region around the $\Gamma$-point of the Brillouin zone of the pristine system and the bands in this region are not well described by an effective mass approximation. We have also computed the local density of states in the vicinity of the charged defects, which is directly accessible in scanning tunneling spectroscopy measurements. Finally, we have studied charged defects in bilayer BP and found that charged adsorbates behave similarly as in the monolayer (but with lower binding energies), while charged intercalants give rise to stronger additional peaks in the local density of states. In conclusion, our calculations provide a detailed understanding of the electronic properties of charged impurities in monolayer and bilayer black phosphorus, which will help realize potential device applications of these materials.

\section{Appendix}

\subsection{Donor Defect States of Monolayer Black Phosphorus}
\begin{figure}
    \centering
    \includegraphics[width=0.5\textwidth]{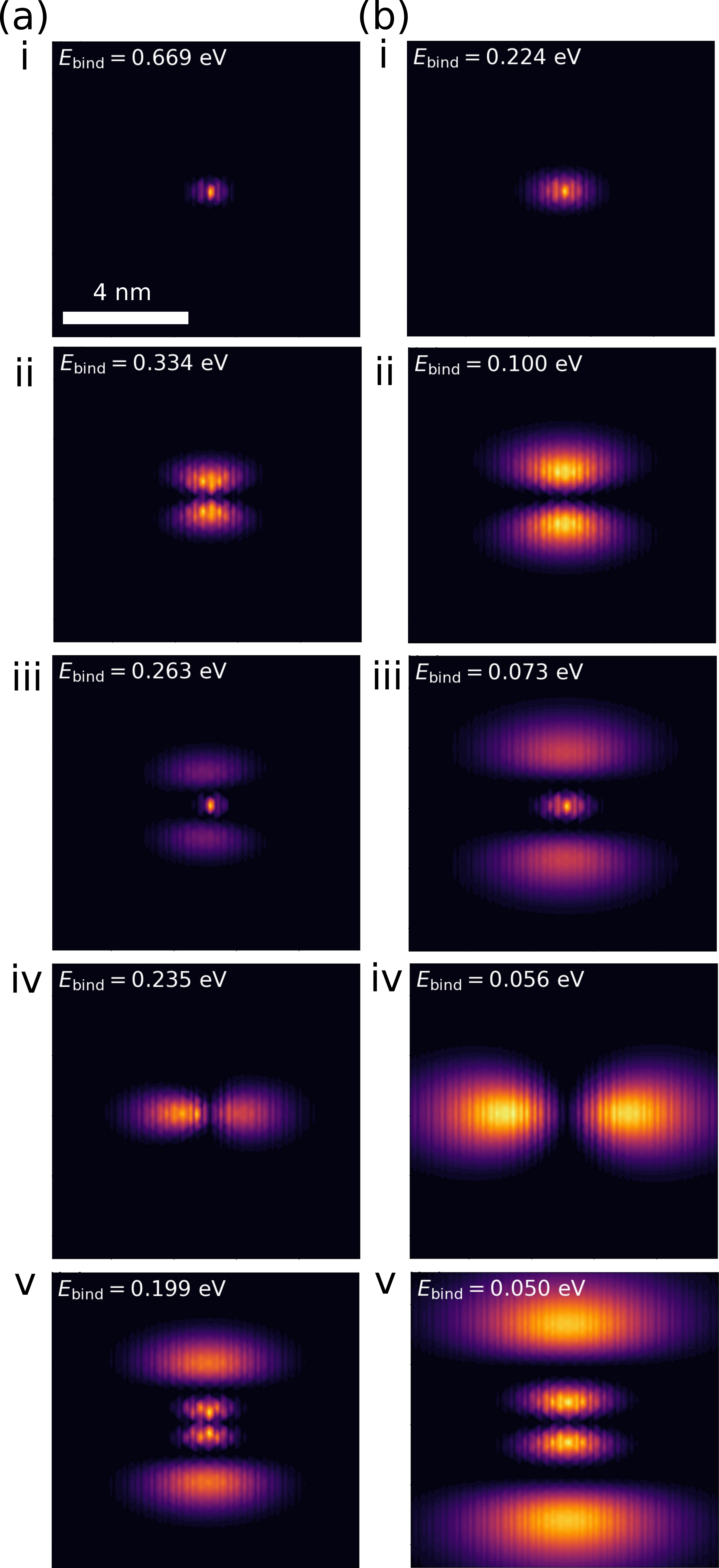}
    \caption{Wavefunctions of the five most strongly bound donor states in monolayer black phosphorus on a dielectric substrate with (a) $\varepsilon_{\text{sub}}=2$ and (b) $\varepsilon_{\text{sub}}=8$, for a defect of height $d=2$~\AA\ above the top sublayer. All defect binding energies $E_{\text{bind}}$ are given relative to the valence band edge.}
    \label{fig:donor_wfc}
\end{figure}

\begin{figure}[b]
    \centering
    \includegraphics[width=0.475\textwidth]{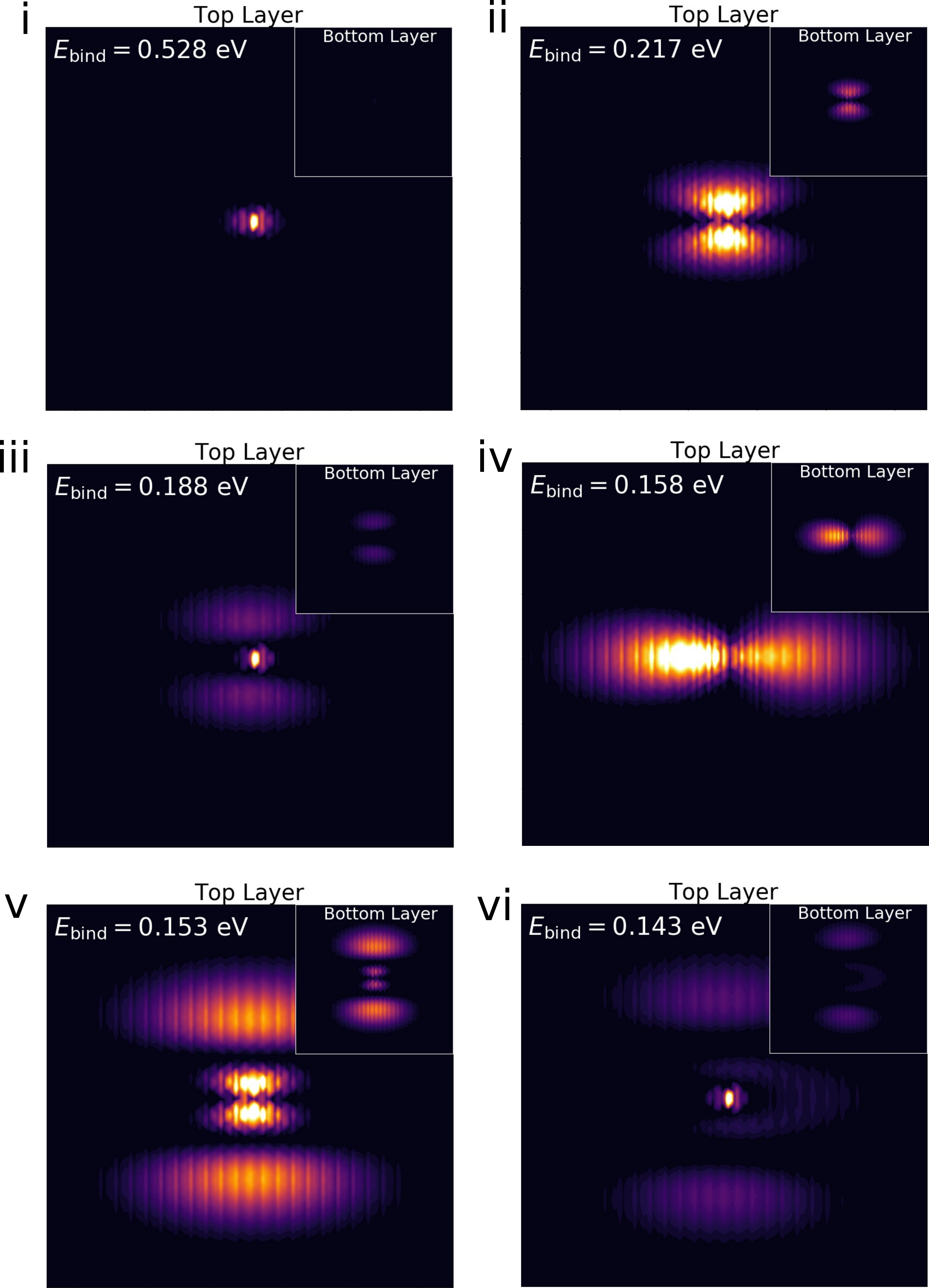}
    \caption{Wavefunctions of the six most strongly bound states of donor adsorbates in bilayer BP in the top monolayer. Results for a positively charged adsorbate with $d=2$~\AA\ above the top sublayer and $\epsilon_{\text{sub}}=1$. The wavefunctions in the bottom monolayer are shown in the inset.}
    \label{fig:bilayer_wfc_donor}
\end{figure}

In Fig.~\ref{fig:donor_wfc} we show the five most strongly bound defect states for a positively charged adsorbate placed $d=2$~\AA\ above the top sublayer of monolayer black phosphorus on a substrate with (a) $\varepsilon_{\text{sub}}=2$ and (b) $\varepsilon_{\text{sub}}=8$.
In Fig. \ref{fig:bilayer_wfc_donor} we show the wavefunctions of the donor adsorbate on freestanding bilayer BP. 

\bibliographystyle{unsrt}
\bibliography{blackphosphorus}
\end{document}